\documentclass[12pt]{iopart}
\usepackage{epsfig}

\begin{document}
\renewcommand{\t}[1]{\mathrm{#1}}            % aufrechte Buchstaben in mathmode
\renewcommand{\vec}[1]{\mathbf{#1}}          % Vector
\renewcommand{\d}[1]{\t{d}#1\,}              % Differential d
\newcommand{\cc}[1]{\overline{#1}}           % komplex konjugiert (durch Überstrich)

\def\la{\langle}
\def\ra{\rangle}

\title{Quantum optical time-of-arrival model in three dimensions}
\author{V. Hannstein*, G. C.~Hegerfeldt* and J. G. Muga\dag} 
\address{* Institut f\"ur Theoretische Physik, Universit\"at G\"ottingen,
Friedrich-Hund-Platz~1, 37077 G\"ottingen, Germany}
\address{\dag Departamento de Qu\'\i{}mica-F\'\i{}sica, UPV-EHU, Apdo. 
644, Bilbao, Spain}
%\date{\today}

\begin{abstract}
We investigate the three-dimensional formulation of a recently proposed
operational arrival-time model. It is shown that within typical
conditions for optical transitions the results of the simple 
one-dimensional version are generally valid. Differences that may occur
are  consequences of Doppler and 
momentum-transfer effects.  Ways to minimize these are discussed.  
\end{abstract}

\maketitle

\section{Introduction}

In quantum mechanics, as opposed to classical physics, the meaning
of arrival time of a particle at a given location is not evident when the
finite extent of the wave function and its spreading become relevant,
such as for  cooled atoms dropping out of a trap. Moreover, in the quantum case
one will expect an arrival-time distribution, and there are different
theoretical proposals for it, see e.g. 
\cite{All:69,Ki:74,We:86,BlJa:96,Gi:97,Le:98,AOPRU:98,Ha:98,KoWo:99,LeJuPiUr:00,Ga:02,Ru:02},
the review in \cite{MuLe:00}, and the book \cite{MuSaEg:02}. These
arrival-time distributions have been controversially discussed in the
literature since they are derived through purely theoretical arguments
rather than operationally, i.e. without specifying a
measurement procedure. But from the experimental point of view this is
of course very important.

How then to measure the arrival time of a single quantum mechanical
particle at a given location when the extent of its wave function
becomes important? Here one has to distinguish between  first arrival
and repeated arrivals as in the case of a pendulum.
For the former, the measurement procedure has to ensure that the arrival is
indeed the first, not the second or third, i.e. that the particle did
not already arrive {\em before}. This implies a sort of continuous or
rapidly repeated observation of the same single particle in
question. Doing the experiment again and again with many identically
prepared  particles will then give an arrival-time distribution. 

To capture these requirements in a model it was 
suggested in \cite{DaEgHeMu:02} to consider a particle (``atom'') with two
internal levels, which are connected by an optical transition, and to
illuminate the arrival region by a laser. When the wave function of
the particle enters the region  the laser will populate the upper level and
one can watch for the first spontaneous photon. The 
detection time of this photon can then be taken as the first arrival
time in this region, at least  approximately.

The model contains of course simplifying idealizations, which are indeed 
quite essential for extracting useful information. In general a 
measurement model should be flexible enough to capture the essence of the 
experiment and still provide, hopefully, close links with 
the schema of fundamental theory. 
That the fluorescence-based model is of such kind has been already
demonstrated: 
the limits or operations by which theoretical time-of-arrival 
distributions, such as the flux or current density and Kijowski's
distribution,  can be obtained by measurement 
have been determined \cite{DaEgHeMu:02,HeSeMu:03}.
For other applications or extensions of the model see
\cite{DaEgHeMu:03,NaEgMuHe:03b,NaEgMuHe:03a,RuDaNaMuHe:04,HeSeMuNa:04}.
It should not be surprising that the conditions required for these  
``perfect measurements'' may be rather
extreme, or complicated to realize. The important point is that at least 
one knows what should be done, and one can reasonably understand and try to 
minimize the deviations from the ideal results.   

A very basic simplifying assumption of all applications of the 
model so far 
has been the restriction of the atomic motion to one dimension
(1D), associated 
with the $x$ axis hereafter,
perpendicular 
to the laser (traveling wave) direction $y$. This approximation could
in principle  be badly wrong because of transversal velocity components   
in the incident atomic wave function or because of momentum transfer 
from the laser. 

The aim of this paper is to generalize the fluorescence 
model to  
an arbitrary three dimensional wave packet impinging on a plane ($x=0$) which 
separates a laser-illuminated region ($x>0$) from a non-illuminated
region ($x<0$). 
It will be shown that possible deviations from the 1D model can be
attributed to an additional detuning term which is of kinetic origin.
This {\em kinetic detuning} consists of a Doppler term and a term which is
due to a momentum transfer from the laser field to the
atom. It will be seen that deviations of the time-of-arrival
distribution from the 1D model can be ascribed to two effects. First,
there is 
increased or decreased reflection for blue or red kinetic detuning,
respectively and, second, there is a less efficient driving of the
atomic transition 
for both signs of the kinetic detuning which leads to a delay in the
distribution. Numerical examples 
will be shown for two typical cases. For oblique incidence the kinetic
effects can be compensated by appropriate laser detuning, whereas for
a wave packet prepared with a very small width in laser direction this
is not possible. The numerical results also show that the 1D model is
valid over a wide range of parameters. 

It will also be shown that the first photon distribution tends in some
limit to the $x$ component of the total flux through the plane $x=0$,
which is the natural generalization of a main result in \cite{DaEgHeMu:02}.

\begin{figure}
\begin{center}
  \epsfig{file=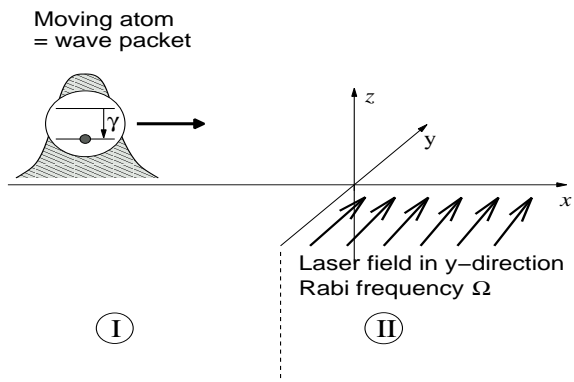, width=7.5cm,height=5cm}
   \caption{\label{F1}Schematic picture of the 
     atom wave packet moving towards the laser-illuminated region.}
\end{center}     
\end{figure}

\section{The 3D model}
\label{3Dm}

As in \cite{DaEgHeMu:02} we consider a two-level atom which moves from
the left towards the
region $x>0$ which is illuminated by a traveling-wave laser directed parallel
to the $y$ axis (see \fref{F1}). The laser frequency may be on resonance or 
be detuned with respect to  the atomic transition.  
We use the dipole and rotating wave approximations and employ the quantum
jump approach \cite{He:93,DaCaMo:92,Ca:93}. In this approach the time
development between two photon detections is given by a nonhermitean
``conditional'' Hamiltonian so that the norm of the wave function
decreases; the norm squared at time $t$ just gives the probability for
no photon detection {\em until} $t$. In  a laser-adapted interaction picture
(i.e. with $H_0 = \hbar \omega_{\t{L}}|2\rangle \langle 2|$) the conditional
Hamiltonian becomes
\begin{equation}
H_{\t{c}}= \frac{\hat{p}_x^2}{2m} + \frac{\hat{p}_y^2}{2m} +
\frac{\hat{p}_z^2}{2m} + V(\hat{\vec{x}})
\end{equation}
with 
\begin{equation}
V(\hat{\vec{x}})=
\frac{\hbar}{2}\Omega\theta(\hat{x})\left[\t{e}^{\rmi
    k_{\t{L}}\hat{y}}|2\rangle 
  \langle 1| + \t{e}^{-\rmi k_{\t{L}}\hat{y}}|1\rangle \langle 2|\right]
- \hbar\left(\Delta_{\t{L}} + \rmi\gamma/2\right)|2\rangle\langle2|,
\end{equation}
where $\gamma$ is the decay (Einstein) constant of the excited state
$|2\rangle$, $\Omega$ is the  
Rabi frequency, $\Delta_{\t{L}}$ the detuning (laser angular frequency minus 
atomic-transition angular frequency), $k_{\t{L}}$ the laser  wavenumber, 
and $\theta$ the Heaviside function. The hats on positions
and momenta ($\hat{\bf{x}},\hat{x},\hat{y},...,\hat{p}_x,...$) 
denote operators to be distinguished from the corresponding 
ordinary vectors or c-numbers. Boldface capital Greek letters will
designate atomic two-component wave functions,  with the components for
excited and ground state denoted by superscripts.      

The operational time-of-arrival distribution $\Pi(t)$ 
for an atomic ensemble represented by a given wave function is defined as the 
temporal distribution of the first photon detected for each atom, and
this is just the  rate of the decrease of probability of  detecting no
photon.  Hence one has the general result
\begin{equation}
\Pi(t) = -\rmd ||\vec{\Psi}(t)||^2/\rmd t~.
\end{equation}
By means of the Schr\"odinger equation and the above form of $H_{\t{c}}$ one
easily finds for the two-level system
\begin{equation}\label{Pi}
\Pi(t) = \gamma\int_{-\infty}^\infty
\,\t{d}^2\vec{x}|\Psi^{(2)}(\vec{x},t)|^2.
\end{equation}
To calculate $\Pi$ it is useful and physically illuminating 
to first obtain the eigenstates of the conditional Hamiltonian, 
\begin{equation}
\label{ev}
H_{\t{c}}\vec{\Phi}_{\vec{k}}=E_{\vec{k}}\vec{\Phi}_{\vec{k}}.
\end{equation}
Since $V(\hat{\vec{x}})$
does not depend on $\hat{z}$, the $z$ component of momentum is conserved 
so that there is free motion in $z$-direction and we 
only have to consider the two-dimensional motion in the $x$-$y$-plane.
Therefore, from now on the position and momentum (wavenumber) vectors
will always be {\em two-dimensional}.

Since the incident atoms are initially in the ground state and far from
the laser region and are moving in the positive
$x$-direction,   
we need the stationary scattering eigenstates which correspond
to ground state plane waves coming in from the left.  
In matrix form, with $|1\rangle \equiv
\textstyle{1\choose0}$ and $|2\rangle
\equiv \textstyle{0\choose 1}$, the (two-dimensional)
conditional Hamiltonian can be written
as
\begin{equation}
H_{\t{c}}=\frac{\hat{p}_x^2}{2m} + \frac{\hat{p}_y^2}{2m}
-\hbar(\Delta_{\t{L}} + \rmi\gamma/2)
\left( \begin{array}{cc} 0 & 0 
\\
0 &
  1\end{array}\right) 
+ \frac{\hbar}{2} \theta(\hat{x})
\left( \begin{array}{cc} 0 &
  \Omega \t{e}^{-\rmi k_{\t{L}}\hat{y}}
\\
\Omega \t{e}^{\rmi k_{\t{L}}\hat{y}} &
  0 \end{array}\right)~.
\end{equation} 
For $x<0$ we  use the ansatz
\begin{equation}
\label{phi1}
\vec{\Phi}_{\vec{k}}^{\t{I}}(\vec{x})=\frac{1}{2\pi}\left( \begin{array}{c}
\t{e}^{\rmi\vec{k}\cdot\vec{x}}+R_1\t{e}^{\rmi\vec{k}^{\prime}\cdot\vec{x}}
\\ R_2\t{e}^{\rmi\vec{q}\cdot\vec{x}}
\end{array} \right)
\end{equation}
with $(2\pi)^{-1}$  included for delta normalization in
two-dimensional $\bf{k}$ space  and with
\begin{equation}\label{q}
E_{\vec{k}}= \frac{\hbar^2\vec{k}^2}{2m}=\frac{\hbar^2\vec{k}^{\prime2}}{2m}
\quad \t{and} \quad \vec{q}^2
=\vec{k}^2+\frac{m}{\hbar} (\rmi\gamma+2\Delta_{\t{L}})~.
\end{equation}
At this point the individual components of $\vec{k}^\prime$ and
$\vec{q}$ are still 
unknown. They will be determined later from the matching conditions at
$x=0$. For the region $x \ge 0$ we use, in analogy to the 1D model
\cite{DaEgHeMu:02,NaEgMuHe:03b}, a plane wave ansatz of the form
\begin{equation}
\vec{\Phi}_{\vec{k}}^{\pm}(\vec{x})=\left(\begin{array}{c}
1 \\ \frac{2\lambda{_{\pm}}}{\Omega}\t{e}^{\rmi k_{\t{L}}y}
\end{array}\right)\t{e}^{\rmi\vec{k}^{\pm}\cdot\vec{x}} \equiv
|E_{\pm}\rangle \t{e}^{\rmi\vec{k}^{\pm}\cdot\vec{x}} 
\label{ansa} 
\end{equation}
with as yet unknown $\lambda{_{\pm}}$ and $\vec{k}^{\pm}$. Inserting
this into the eigenvalue equation (\ref{ev}) 
the exponential
$\t{e}^{\rmi\vec{k}^{\pm}\cdot\vec{x}}$ drops out and we get the
matrix equation
\begin{equation}
\label{mateq}
\left(\begin{array}{cc}\frac{\hbar^2}{2m}\vec{k}^{\pm 2} &
    \frac{\hbar}{2}\Omega\t{e}^{-\rmi k_{\t{L}}y}\\ 
\frac{\hbar}{2}\Omega\t{e}^{-\rmi k_{\t{L}}y} &
\frac{\hbar^2}{2m}\vec{k}^{\pm 2}-\hbar(\Delta+\rmi\gamma/2)\end{array}\right)
|E_\pm\rangle = E_{\vec{k}}|E_\pm\rangle~,
\end{equation}
where  
\begin{equation}
\label{gendet} 
\Delta \equiv \Delta_{\t{L}}-\frac{\hbar}{2m}(2k_y^{\pm}k_{\t{L}}+k_{\t{L}}^2) 
\end{equation}
can be regarded as an effective detuning. We shall later analyze 
in detail the 
physical meaning of the second term, which is clearly of kinematic
nature.   
\Eref{mateq} is fulfilled if $\lambda_{\pm}$ and
$\vec{k}^{\pm}$ satisfy
\begin{equation}
\label{lpm}
\lambda_{\pm}=-\frac{1}{4}\left(\rmi\gamma+ 2\Delta\right) \pm
  \frac{\rmi}{4}\sqrt{\left(\gamma-2\rmi\Delta\right)^2 - 4\Omega^2}
\end{equation}
and
\begin{equation}
\vec{k}^{\pm 2}=\vec{k}^2- \frac{2m}{\hbar}\lambda_{\pm}~. 
\end{equation}
Note that $\lambda_{\pm}$ still contains the unknown components $k_x^\pm$ and
$k_y^\pm$. The eigenstate for $x\ge0$ is now a superposition of the form
\begin{equation}
\label{phi2}
\vec{\Phi}_{\vec{k}}^{\t{II}}(\vec{x}) = \frac{1}{2\pi}\left[ C_+ |E_+\rangle
  \t{e}^{\rmi\vec{k}^+\cdot\vec{x}} + C_-|E_-\rangle
  \t{e}^{\rmi\vec{k}^-\cdot\vec{x}} \right].
\end{equation}
The coefficients $C_\pm$ will be determined from the  matching conditions.

\subsection{Matching Conditions}

At $x=0$ the usual matching conditions, i.e., 
\begin{equation}
\eqalign{\vec{\Phi}_{\vec{k}}^{\t{I}}(x=0,y)
  &=\vec{\Phi}_{\vec{k}}^{\t{II}}(x=0,y)  
\\
\vec{\Phi}_{\vec{k}}^{\t{I}\,^\prime}(x=0,y) & =\vec{\Phi}_{\vec{k}}^{\t{II}\,^\prime}(x=0,y)} 
\end{equation}
are imposed. This yields the equations
\numparts
\label{match}
\begin{eqnarray}
\label{match1}
\t{e}^{\rmi k_y y }+R_1\t{e}^{\rmi k_y^\prime y}  = 
C_+\t{e}^{\rmi k_y^+ y}+ C_- \t{e}^{\rmi k_y^-
y}
\\
\label{match2}
R_2\t{e}^{\rmi q_yy} = 
\frac{2\lambda_+}{\Omega}C_+\t{e}^{\rmi
(k_{\t{L}}+k_y^+)y} +
\frac{2\lambda_-}{\Omega}C_-\t{e}^{\rmi (k_{\t{L}}+k_y^-)y}
\\
\vec{k}\t{e}^{\rmi k_yy} + \vec{k}^\prime R_1 \t{e}^{\rmi k_y^\prime
  y} =
\vec{k}^+C_+\t{e}^{\rmi k_y^+y} + \vec{k}^-C_-\t{e}^{\rmi k_y^-y}
\\
\vec{q}R_2 \t{e}^{\rmi q_y y} = 
\vec{k}^+\frac{2\lambda_+}{\Omega}C_+\t{e}^{\rmi (k_{\t{L}}+k_y^+)y}
+ \vec{k}^-\frac{2\lambda_-}{\Omega}C_-\t{e}^{\rmi
(k_{\t{L}}+k_y^-)y} 
\end{eqnarray}
\endnumparts
As (\ref{match1}) has to hold for all $y$ one obtains
\begin{equation}
k_y=k_y^{\prime}=k_y^+=k_y^-~,
\label{kkkk}
\end{equation}
and with this one finds from (\ref{q})
  for 
 $k_x^{\prime}$  of the reflected component
\[
k_x^{\prime}=-k_x. 
\]
From (\ref{kkkk}) it follows that 
$\Delta$ is independent of $\vec{k}^\pm$, and consequences of this will be
discussed in more detail in the next subsection. For the
$x$ components of the wave vector inside the laser region one then obtains
\[
k_x^{\pm 2} = k_x^2+\frac{m}{2\hbar}\left(\rmi\gamma+ 2\Delta\right) \mp
\frac{\rmi m}{2\hbar}\sqrt{\left(\gamma- 2\rmi\Delta\right)^2
-4\Omega^2},
\]  
and $k_x^\pm$ are given by the roots of this with positive imaginary part 
so that the wave decays in the laser region. 
Both depend on $k_y$ through the effective detuning $\Delta$.
From (\ref{match2}) one gets by a similar argument 
\[
q_y=k_y+k_{\t{L}},
\]
where $\hbar k_{\t{L}}$ is the momentum transfer from the laser field to the
excited state of the atom. This and (\ref{q}) in turn lead to 
\[
q_x^2=k_x^2+\frac{m}{\hbar}\left(\rmi\gamma + 2\Delta\right) 
\]
for the reflected part of the excited state, and $q_x$ is obtained 
by again taking the root with positive imaginary part so that the excited wave 
decays at long distances from the laser. With these
simplifications the matching conditions (16) now take the
form
\numparts
\begin{eqnarray}
1+R_1 = C_+ + C_-  \\
R_2 = \frac{2\lambda_+}{\Omega}C_++ \frac{2\lambda_-}{\Omega}C_- \\
k_x (1 - R_1)  =
k_x^+C_+ + {k}_x^-C_- \\
q_x R_2  =
k_x^+\frac{2\lambda_+}{\Omega}C_+ + k_x^-\frac{2\lambda_-}{\Omega}C_- 
\end{eqnarray}
\endnumparts
which look formally just same as in the 1D case, only with $k^\pm$ and $q$
replaced by $k^\pm_x$ and $q_x$ and with $\lambda_\pm$ defined
as in (\ref{lpm}). Accordingly the coefficients $R_1$, $R_2$, $C_{\pm}$ have
the same form as in the 1D model,
\numparts
\begin{eqnarray}
C_+ = -2k_x(q_x+k_x^-)\lambda_-/D \\
C_- = 2k_x(q_x+k_x^+)\lambda_+/D \\
R_2 = k_x(k_x^--k_x^+)\Omega/D\\
R_1 = 
[\lambda_+(q_x+k_x^+)(k_x-k_x^-) 
%\\ & 
-\lambda_-(q_x+k_x^-)(k_x-k_x^+)]/D
\end{eqnarray}
and 
\begin{equation}
D= \lambda_+(k_x+k_x^-)(q_x+k_x^+)-\lambda_-(k_x+k_x^+)(q_x+k_x^-).
\end{equation}
\endnumparts
Finally, taking into account the matching conditions, (\ref{phi1}) and
(\ref{phi2}) take
the form
\begin{equation}
\vec{\Phi}_{\vec{k}}^{\t{I}}(\vec{x})=\frac{1}{2\pi}\left( \begin{array}{c}
\left[\t{e}^{\rmi k_xx}+R_1\t{e}^{-\rmi k_xx}\right]\t{e}^{\rmi k_y y}
\\ R_2\t{e}^{-\rmi q_xx}\t{e}^{\rmi(k_y+k_{\t{L}})y}
\end{array} \right)
\label{wfI}
\end{equation}
for $x<0$ and 
\begin{equation}
\vec{\Phi}_{\vec{k}}^{\t{II}}(\vec{x}) = \frac{1}{2\pi}\left[ C_+
\left(
\begin{array}{c}
1
\\
\frac{2\lambda_+}{\Omega}\t{e}^{\rmi k_{\t{L}}y}
\end{array}
\right)
  \t{e}^{\rmi k_x^+x} 
+ C_-
\left(
\begin{array}{c}
1
\\
\frac{2\lambda_-}{\Omega}\t{e}^{\rmi k_{\t{L}}y}
\end{array}
\right)
  \t{e}^{\rmi k_x^-x} \right]\t{e}^{\rmi k_y y}
\label{wfII}
\end{equation}
for $x> 0$.

\subsection{Effective detuning: Doppler shift and momentum transfer}

Using (\ref{kkkk}), the effective detuning 
(\ref{gendet}) can be  
written as  
\begin{equation}
\Delta = \Delta_{\t{L}}-\Delta_{\t{K}},
\label{deldel}
\end{equation}
with the {\em kinetic detuning}
\[
\Delta_{\t{K}}\equiv\frac{\hbar}{2m}(2k_yk_{\t{L}}+k_{\t{L}}^2)=
\frac{\hbar}{2m}(k_y + k_{\t{L}})^2 - \frac{\hbar}{2m}k_y^2
\]
so that $\hbar \Delta_{\t{K}}$ is the $y$ momentum component
contribution   to the kinetic energy difference   
between ground state and excited state, as seen from  the terms $k_y$ 
and $k_y+k_{\t{L}}$ in the plane wave exponents 
in (\ref{wfI}) and (\ref{wfII}).
We may also express $\Delta_{\t{K}}$ in terms of the speed of
light in vacuum $c$, the laser
angular frequency $\omega_{\t{L}}=k_{\t{L}} c$, the $y$ component of the 
incident velocity $v_y=\hbar k_y/m$, and of a velocity
$v_{\t{R}}\equiv \hbar k_{\t{L}}/m$, namely
\begin{equation}
\label{deld}
\Delta_{\t{K}}=\frac{\omega_{\t{L}}}{c}\left(v_y+\frac{v_{\t{R}}}{2}\right). 
\end{equation}

The form of (\ref{deld})   
can also be derived  
by classically modeling the resonance 
condition with simple energy and momentum conservation arguments. To
see this we rewrite it as 
\begin{equation}
\label{deld1}
\Delta_{\t{K}}=\omega_{\t{D}} + \omega_{\t{R}}
\end{equation}
with $\omega_{\t{D}}\equiv\omega_{\t{L}} v_y/c$ and
$\omega_{\t{R}}\equiv\hbar
k_{\t{L}}^2/2m=\omega_{\t{L}}v_{\t{R}}/2c$. The first term is the
ordinary Doppler shift due to the incident velocity component $v_y$
and the second can be viewed as a frequency shift caused by the change
of the atomic kinetic energy due to the absorption of a laser
photon. Recall that the momentum transfer from the laser is only
directed along the $y$ direction. It is interesting to note that the
kinetic detuning in (\ref{deld1}) has in fact
already the form of a Doppler shift formula with the role of the
velocity played by the average of the velocity components for ground state
and excited state along the direction of the laser. Indeed, for fixed
non-zero values of $k_y$ and $k_{\t{L}}$ in a stationary wave
the two effects cannot be disentangled and act simultaneously in the 
form of a single kinetic detuning.
It is, however, possible to distinguish their consequences  
for wave packets, as discussed below, since $k_{\t{L}}$ is fixed by the 
laser wavelength, whereas $k_y$ varies for each incident 
atomic momentum 
component, or by sweeping over the incidence angle.

Note also the 
signs in  the effective total detuning (\ref{deldel}):    
a large, positive, kinetic detuning (energy transfer) implies a red
shift whereas a large laser, positive detuning amounts to a blue
shift. They may compensate each other, as we shall illustrate
afterwards, while $\omega_{\t{D}}$ and $\omega_{\t{R}}$ may also
cancel each other for a negative velocity $v_y$ opposed to the laser beam.  

\section{The First-Photon Distribution}
\label{results}

Until detection of the first photon,  the conditional time
development of a wave 
packet $\vec{\Psi}(\vec{x},t)$ corresponding to a particle in the 
internal ground state coming in from the left can now be written as in
the 1D case in terms of the plane-wave solutions,
\begin{equation}
\vec{\Psi}(\vec{x},t)=\int_0^\infty
\!\d{k_x}\!\int_{-\infty}^\infty\!\d{k_y}\tilde{\psi}(\vec{k})
\vec{\Phi_{\vec{k}}}(\vec{x}) \t{e}^{-\rmi\hbar\vec{k}^2t/2m},
\label{psixt}
\end{equation}
where $\tilde{\psi}(\vec{k})$ is the momentum amplitude the wave
packet would have as a freely moving particle at $t=0$. Inserting this
into (\ref{Pi}) one obtains the first-photon distribution. This gives
a six-dimensional integral of which the $x$, $y$, and
$k_y^\prime$ integrations can be
carried out analytically, yielding
\begin{eqnarray}
\fl \Pi(t) = \frac{\gamma}{2\pi}\int_0^\infty
\!\!\!\d{k_x} \!\!\int_0^\infty
\!\!\!\d{k_x^\prime}\!\!\int_{-\infty}^\infty\!\!\!\d{k_y}
\cc{\widetilde{\psi}(k_x,k_y)} \widetilde{\psi}(k_x^\prime,k_y)
\left[
  \frac{\rmi \cc{R_2}R_2^\prime}{q_x^\prime-\cc{q}_x} 
+
\rmi\frac{4\cc{C_+}C_+^\prime\cc{\lambda}_+\lambda_+^\prime}{\Omega^2(k_x^{+\prime}
  - \cc{k_x^{+}})} 
\right. \\ \left. {}
+
  \rmi\frac{4\cc{C_+}C_-^\prime\cc{\lambda}_+\lambda_-^\prime}{\Omega^2(k_x^{-\prime} - \cc{k_x^{+}})}%}
 + \rmi\frac{4\cc{C_-}C_+^\prime\cc{\lambda}_ - \lambda_+^
    \prime} {\Omega^2 (k_x^{+\prime} - \cc{k_x^{-}})} +
  \rmi\frac{4\cc{C_-}C_-^\prime\cc{\lambda}_-\lambda_-^\prime}{\Omega^2(k_x^{-\prime} - \cc{k_x^{-}})}\right] \t{e}^{-\rmi\hbar(k_x^{\prime 2}-k_x^2)t/2m}. \nonumber
\end{eqnarray}
This form of the first-photon distribution will be now used to 
numerically look for deviations from the 1D results 
in different parameter regimes.   

The relevance of possible deviations can be estimated by 
comparing the two-dimensional (2D) eigenstates with those of the
1D model which are given by \cite{DaEgHeMu:02,NaEgMuHe:03b}
\begin{equation}
\fl \vec{\Phi}^{\t{(1D)}}_{{k}}({x}) = 
    \cases {\frac{1}{\sqrt{2\pi}} \left( \begin{array}{c}
          \t{e}^{\rmi k x}+R_1
          \t{e}^{-\rmi kx} \\
          R_2\t{e}^{-\rmi q x}
        \end{array}\right)  &for $x < 0$ \\
  \frac{1}{\sqrt{2\pi}} \left[ C_+ \left(\begin{array}{c}
        1 \\ \frac{2\lambda_+^{\t{(1D)}}}{\Omega}
      \end{array}\right)
    \t{e}^{\rmi k_+x}   
    + C_-\left(\begin{array}{c}
        1 \\ \frac{2\lambda_-^{\t{(1D)}}}{\Omega} \end{array}\right)
    \t{e}^{\rmi k_-x} \right] & for   $ x\ge 0$} 
\end{equation}
where 
\begin{eqnarray*}
q =\sqrt{k^2+\frac{2m}{\hbar}(\Delta_{\t{L}}+\rmi\gamma/2)}~,
\quad k_\pm = k^2-
\frac{2m}{\hbar}\lambda_\pm~,
\\ \lambda_\pm =-
\frac{1}{4}\left(\rmi\gamma+2\Delta_{\t{L}}\right) \pm
\frac{\rmi}{4}\sqrt{(\gamma-2\rmi\Delta_{\t{L}})^2-4\Omega^2} 
\end{eqnarray*}
and roots are taken with positive imaginary parts.  
One can go from 2D to 1D formulas 
($q_x \rightarrow q$, 
$\lambda^{(\t{2D})}_\pm \rightarrow \lambda^{(\t{1D})}_\pm$,
$k^\pm_x\rightarrow k_\pm$)
simply by dropping the \emph{kinetic detuning}  
$\Delta_{\t{K}}$, which is thus identified as the 
\emph{physical reason} for possible deviations from the 1D results. 

For optical transitions, typical values of $\Omega$, $\gamma$ or
$\Delta_{\t{L}}$ are of order $10^{7}$ s$^{-1}$, while the shift
$\omega_{\t{R}}$ in (\ref{deld}) for Cs is of order 
$10^{5}$ s$^{-1}$. For perpendicular incidence, i.e. $\la k_y\ra=0$,
and for velocities $v_y$ corresponding to a $\Delta_y=1 \mu$m wave packet, the 
Doppler shift $\omega_\t{D}$ is even smaller, 
namely of order $10^3$ s$^{-1}$. So the 1D model results may be expected 
to hold quite generally in a broad (ordinary) range of parameters.      
The number of parameters is rather large so a full systematic analysis 
of all possible cases in the vast parameter space is out of the question.
Nevertheless, it is worth examining the convergence from 2D to 1D
results in some typical cases. Since the laser and kinetic  
detunings,  $\Delta_{\t{L}}$ and $\Delta_{\t{K}}$, are always combined
together in $\lambda_\pm$, 
$k_x^\pm$ and $q$, one possible limit for pure 1D behaviour is 
$|\Delta_{\t{L}}|\gg|\Delta_{\t{K}}|$. Moreover, from 
\eref{deldel} it is also possible to find 1D results if the 
two contributions to the effective detuning cancel each other, i.e. if
$\Delta_{\t{L}}=\Delta_{\t{K}}$, as we shall discuss below. 
$\Delta_{\t{K}}$ may also be zero because of a cancellation between 
Doppler and momentum-transfer shifts in oblique incidence.    
One may similarly reduce the equations to 1D if  
$\Omega$ or  $\gamma$ are large with respect to all other frequency 
parameters.  
For strong driving, $\Omega\gg\gamma,\Delta_{\t{L}}$, more detailed 
sufficient conditions would be   
$k_y \ll k_x^2/k_{\t{L}}$, $k_y,k_{\t{L}} \ll
\frac{m}{\hbar}\Omega/k_{\t{L}}$, and $k_{\t{L}}^2 \ll k_x^2$.  
This means that a two-dimensional wave packet, with given mean
momenta $\langle k_x\rangle$, $\langle k_y\rangle$
and momentum widths $\Delta_{k_x}$,
$\Delta_{k_y}$, can be described by the 1D model if it has only small
$k_y$ components and if at the same time $k_{\t{L}}$ is
small compared to the
$k_x$ component and $\Delta_{\t{K}}\ll\Omega$. 
In particular the momentum spread of the wave-packet
in $y$ direction should be small compared to the momentum spread in
$x$ direction. This, on the other hand, corresponds to a large width
$\Delta_y$ in real space which is very satisfactory since for
$\Delta_y \rightarrow \infty$ the two-dimensional wave packet tends to
a one dimensional one. 

\begin{figure}
  \begin{center}
    \epsfig{file=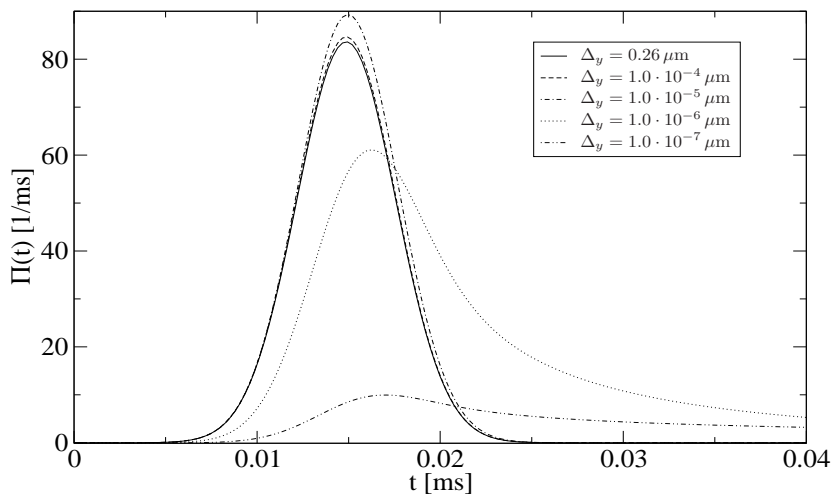,width=11cm,height=6.5cm}
    \caption{\label{F2} Arrival time density for a caesium atom with the
      parameters $\Omega=1.67 \cdot 10^8 \,\t{s}^{-1}$,
      $\gamma=3.3\cdot 10^8 \,\t{s}^{-1}$, prepared at $t=0$ as a
      minimum-uncertainty product Gaussian wave packet with
      $\Delta_x=0.24\cdot 10^{-6}\,\t{m}$, $\la k_x\ra=0$, and with
      initial average position $\langle x_0\rangle=-1.32\cdot 10^{-6}
      \,\t{m}$, and velocity $\la v_0\ra =9\,\t{cm/s}$. The solid line
      distribution coincides with the 1D distribution.}
  \end{center}
\end{figure}

\begin{figure}
  \begin{center}
    \epsfig{file=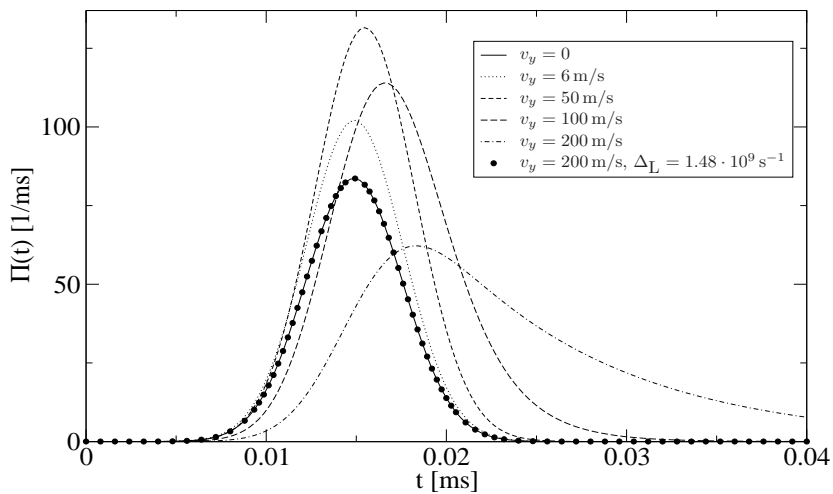,width=11cm,height=6.5cm}
    \caption{\label{F3} Arrival time density for Gaussian wave packets with
      different positive initial average velocities
      $\langle v_y\rangle$. Atomic parameters as in \fref{F2}. The
      dotted graph shows a distribution with
      $\Delta_{\t{L}}=\Delta_{\t{K}}$ for compensation of the kinetic effects.}
  \end{center}     
\end{figure}

\begin{figure}
  \begin{center}
    \epsfig{file=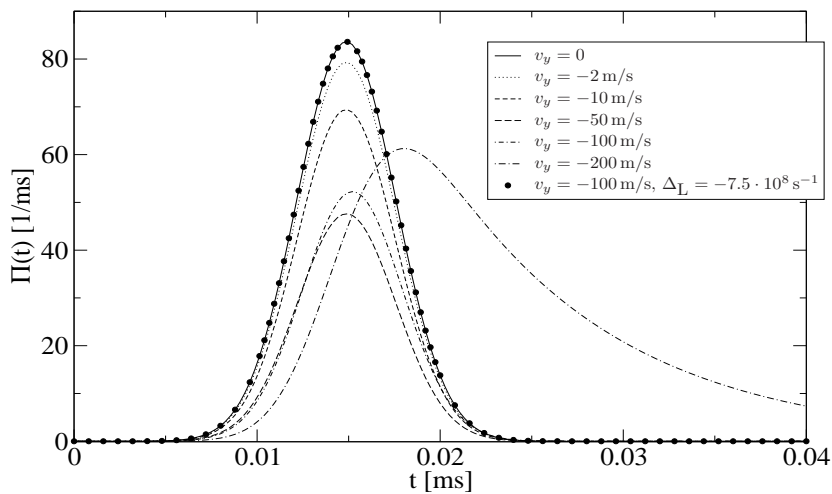, width=11cm,height=6.5cm}
    \caption{\label{F4} Arrival time distribution for wave packets with
      different negative initial average velocities
      $\langle v_y\rangle$. Atomic parameters as in \fref{F2}. The
      dotted graph again shows a distribution with
      $\Delta_{\t{L}}=\Delta_{\t{K}}$ for compensation of the kinetic effects.}
  \end{center}     
\end{figure}

This behaviour can be seen in \fref{F2}
for arrival time distributions of minimum-uncertainty product Gaussian
wave packets with different widths $\Delta_y$ and fixed width
$\Delta_x$ at the preparation time $t=0$  (see the caption for
details), where parameters of Cs have been used. For
large widths $\Delta_y$ the time of arrival distribution is identical
to the one which is obtained from a one-dimensional Gaussian wave packet
of width $\Delta_x$ in the 1D model. For very small widths
$\Delta_y$ first a slight enhancement in the height of the
distribution and then a delay in the form of an enhanced tail of the
distribution can be seen. 
The physical reasons for these two effects are a dependence of the
reflection coefficient on the velocity in $y$ direction and the
Doppler effect, respectively, as will be explained in the next paragraph.
It is seen in \fref{F2} that a 
significant deviation from the 1D distribution occurs for the Cs
parameters used here if the wave
packet is prepared with a width at least three orders of magnitude
smaller in $y$ direction than in $x$ direction.
A deviation from the 1D time of arrival can also be seen if a sufficiently
large momentum $\la k_y\ra$ is chosen, which corresponds to oblique 
incidence of the atoms onto the laser region. \Fref{F3} shows
numerical results for Cs wave packets with different positive average
initial momenta. Again there are two effects: first an \emph{increase}
in height of the distribution and then for even larger momenta a delay. 
For negative initial average momenta (see \fref{F4}) on the other
hand one first sees a \emph{decrease} in height of the distributions and then
also a delay. 

\begin{figure}
  \begin{center}
    \epsfig{file=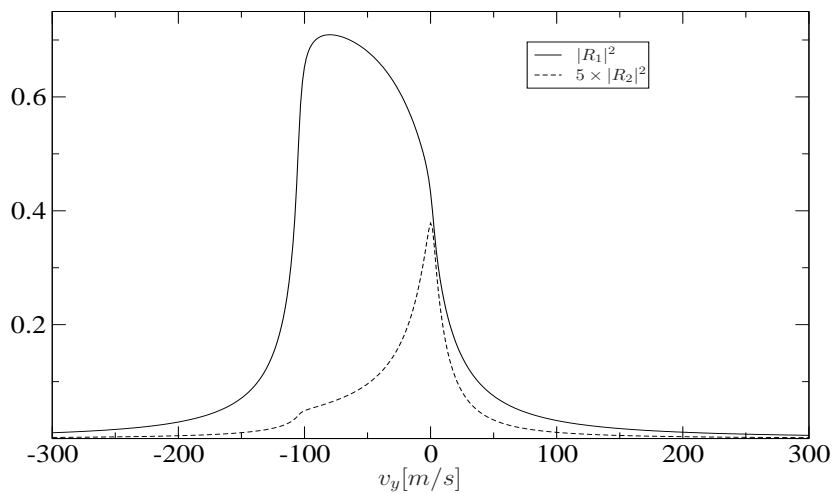, width=11cm,height=6.5cm}
    \caption{\label{F5} Plot of the modulus squared of the reflection
      coefficients $R_1$ and $R_2$ as a function of the incident
      velocity $v_y$. Parameter values as in \fref{F2}. The graph
      of $|R_2|^2$ is stretched by a factor of 5.}
  \end{center}   
\end{figure}

The change in height of the distributions is due to
the $k_y$ dependence of the reflection coefficients, as will be
explained now. Compared to the
eigenstate with $k_y=0$ the eigenfunctions with positive $k_y$ have a
smaller reflected part whereas the eigenfunctions with negative $k_y$
have a larger reflected part, as shown in \fref{F5}. This
behaviour can be understood via the dipole force and the effective
detuning of \eref{deldel} as follows. Positive values of $k_y$ lead to an
effective red detuning while negative values lead to blue
detuning. Therefore states with positive $k_y$ are pushed into the
laser region resulting in decreased reflection and states with
negative $k_y$ are pulled out of the laser region resulting in
increased reflection. The delay is due to the diminished efficiency of
the laser because of the Doppler shift which also explains the decrease in
reflection for very high negative momenta seen in \fref{F5}. 
For the parameter values of Caesium used here one again has to insert
rather extreme values of $\langle k_y\rangle$, i.e. at least three
orders of magnitude larger than $\langle k_x\rangle$, to see
deviations from the 1D distribution. The effects of nonzero initial
average momentum $k_y$ can be compensated by choosing a laser detuning
$\Delta_{\t{L}} = \Delta_{\t{K}}$ which leads to zero average
effective detuning. Since quantum effects in the time-of-arrival can
only be observed for sufficiently small incident velocities and since
the 3D distribution corresponds to the 1D distribution with $v_x$ as
incident velocity the oblique incidence could be useful in order to
obtain broader distributions.

The effects of very
small widths $\Delta_y$, seen in \fref{F2}, can also be understood
by means of the two effects explained above. The delay is again due to the Doppler effect. Indeed, since small uncertainty in position means
large uncertainty in momentum, a wave packet prepared with small $\Delta_y$
will broaden very rapidly so that the tails of the wave packet move away
from the center. Due to the Doppler effect these tails see a detuned
laser and the wave packet as a whole is excited less efficiently. Since
there is a shift with positive and negative sign for different 
components, it is not possible to compensate this effect by laser
detuning. The increase in height of the distribution is a little bit
subtler. Starting from $k_y=0$ the decrease of $|R_1|^2(k_y)$ in
\fref{F5} when going to positive values of $k_y$ is slightly steeper
than the increase in the other direction. 
Therefore there is less reflection for increasing $\Delta_{k_y}$,
although both positive and negative values of $k_y$ occur with the
same weight in the wave packet.

Deviations from the 1D distribution also occur for a large
shift $\omega_{\t{R}}$, i.e. if $k_{\t{L}}^2$ is large compared to the
other relevant parameters. However, this requires an incident
velocity smaller than the recoil velocity $v_{\t{R}}$ and, for a laser
transition in the optical range, a metastable transition with a
lifetime of the order of one tenth of a second.

\section{Deconvolution}
\label{deconvolution}

The operational model to quantum arrival times presented in
\cite{DaEgHeMu:02} and generalized here to three dimensions 
shows that there may arise two potential problems when comparing
operational with ideal distributions, namely 
reflection due to the laser field and detection delay. Reflection in
the ground state  means that no photon is emitted and hence the arrival
of the atom is not detected.
Delay is due to the fact that excitation and de-excitation of
the atom take a finite time. If one tries to get rid
of the delay by increasing both the decay rate $\gamma$ and the Rabi frequency
$\Omega$ this will increase reflection in the ground state and thus
lead to a detection decrease. To circumvent this one may try to use a gentle
(i.e., not reflecting) weak excitation and then subtract the long delays in a
suitable manner.  In  \cite{DaEgHeMu:02} this was done by a
deconvolution with the first-photon distribution of an atom at rest as follows.
It was assumed that the ``experimental'' time-of-arrival distribution
$\Pi(t)$ is given by a convolution of a hypothetical ideal
distribution $\Pi_{\t{id}}$ with the (known) first-photon distribution
$W(t)$ of a two-level atom at rest,
\begin{equation}
\Pi = \Pi_{\t{id}}\star W.
\end{equation}
In the limit of large $\gamma$ it was then seen that the resulting
ideal distribution tends to the quantum mechanical flux 
\[
\Pi_{\t{id}}\rightarrow J_\psi
= \frac{\hbar}{2m\rmi}
\left\{\cc{\psi(0,t)}\psi^\prime(0,t)-\cc{\psi^\prime(0,t)}\psi(0,t)\right\}.
\]
A similar calculation can be  carried out here and it
yields the $x$ component of the total flux through the plane $x = 0$,
\[
\Pi_{\t{id}}\rightarrow \cc{J}_x(0,t) = \left[\int_{-\infty}^\infty\d{y}J_x(x,y,t)\right]_{x=0} 
\]
with 
\[
J_x(x,y,t) =
\frac{\hbar}{2m\rmi}
\left\{\cc{\psi(x,y,t)}\frac{\d{\psi}}{\d{x}}(x,y,t)-\cc{\frac{\d{\psi}}{\d{x}}(x,y,t)}\psi(x,y,t)\right\}.
\]
This is the result one would expect as the natural generalization when
carrying over the 1D case to three dimensions.

\section{Conclusions}

We have generalized the quantum optical time-of-arrival model of
\cite{DaEgHeMu:02} to three dimensional space in order to describe 
a more realistic setup and verify the approximations made
there. When considering arrivals at the plane $x = 0$ we have seen that 
only $x$ and $y$ directions have to be considered, which leads to an
effective 2D model.  
Deviations from the 1D model can be described via a kinetic detuning
$\Delta_{\t{K}}$ which consists of two terms for two distinct
effects. One is a detuning of the atomic transition from the laser
frequency due to the Doppler effect for velocity components $v_y$ in the
laser direction. The other is the gain of momentum of the atom in the  
positive-$y$ (laser) direction by absorption of a laser photon.  
Again via the Doppler effect, this leads to a shift in
frequency of the internal transition of the atom in the lab frame.   
A nonzero kinetic detuning in turn leads to two different effects
affecting the time-of-arrival distribution. Blue and red detuning
results in increased and decreased reflection, respectively, and
therefore in a smaller or 
larger height of the distribution. At the same time both detunings
lead to a less efficient driving of the atomic
transition by the laser and thus to a delay in the distribution.   

We have described typical situations for which deviations from the
1D model occur. One is the preparation of a wave packet with very
small width $\Delta_y$. Then the corresponding large momentum width
leads to large transversal velocity components. Another possibility
is oblique incidence with large (positive or negative) mean transversal
momentum $\langle k_y \rangle$. In the latter case the deviations can
be compensated by an appropriate detuning of the laser whereas in the
first case this is not possible.
Inserting realistic atomic parameters it has been shown that over a
wide range of parameters the 1D model is generally applicable. 

A main theoretical result of reference \cite{DaEgHeMu:02} was that the
first-photon distribution tends in some limit to the quantum
mechanical probability current, opening the way towards a measurement of this
quantity. A similar calculation in the generalized
case gives the $x$ component of the total flux through the plane
$x=0$, which is the natural generalization of the 1D result.

\ack{
JGM acknowledges 
``Ministerio de Ciencia y Tecnolog\'\i a-FEDER''
(BFM2003-01003 and ``Acci\'on Integrada''),  and 
UPV-EHU (Grant 00039.310-13507/2001).}

\end{document}